\documentclass{article}

\usepackage{amsmath}
\usepackage{amssymb}
\usepackage{amsfonts}
\usepackage{graphicx}
\usepackage{cite}
\usepackage{color}
\usepackage{enumerate}
\usepackage{float}
\usepackage{authblk}
\usepackage{subcaption}
\usepackage{caption}
\usepackage{hyperref}
\usepackage[dvipsnames,svgnames]{xcolor}
\usepackage{multirow}
\usepackage{array}
\usepackage{lineno}
\usepackage[a4paper, total={6.5in, 10in}]{geometry}
\usepackage{url}

\usepackage{rotating}

\usepackage{longtable}

\begin{document}

\title{Integrating sentiment and social structure to determine
  preference alignments: The Irish Marriage Referendum}


\author[1,*]{David J.P. O'Sullivan} \author[2]{Guillermo
  Gardu\~no-Hern\'andez} \author[1]{James P. Gleeson}
\author[3,$\dagger$]{Mariano Beguerisse-D\'iaz}
\affil[1]{{\footnotesize MACSI, Department of Mathematics and
    Statistics, University of Limerick, Limerick, Ireland}}
\affil[2]{{\footnotesize Sinnia, Mexico City, Mexico}}
\affil[3]{{\footnotesize Mathematical Institute,
    University~of~Oxford,~Oxford~OX2~6GG,~UK}}
\affil[*]{{\footnotesize \tt David.OSullivan@ul.ie }}
\affil[$\dagger$]{{\footnotesize \tt beguerisse@maths.ox.ac.uk}}
\date{\today}

\maketitle

\begin{abstract}
  We examine the relationship between social structure and sentiment
  through the analysis of a large collection of tweets about the Irish
  Marriage Referendum of 2015. We obtain the sentiment of every tweet
  with the hashtags {\tt \#marref} and {\tt \#marriageref} that was
  posted in the days leading to the referendum, and construct networks
  to aggregate sentiment and use it to study the interactions among
  users. Our results show that the sentiment of mention tweets posted
  by users is correlated with the sentiment of received mentions, and
  there are significantly more connections between users with similar
  sentiment scores than among users with opposite scores in the
  mention and follower networks. We combine the community structure of
  the two networks with the activity level of the users and sentiment
  scores to find groups of users who support voting `yes' or `no' in
  the referendum. There were numerous conversations between users on
  opposing sides of the debate in the absence of follower connections,
  which suggests that there were efforts by some users to establish
  dialogue and debate across ideological divisions. Our analysis shows
  that social structure can be integrated successfully with sentiment
  to analyse and understand the disposition of social media users.
  These results have potential applications in the integration of data
  and meta-data to study opinion dynamics, public opinion modelling,
  and polling.
\end{abstract}


\section{Introduction}\label{sec:intro}

The Republic of Ireland held a referendum to legalise same-sex
marriage on the $22$ of May $2015$. This referendum saw a high turnout
(60.52\% of voters), and the final result was a 62\% majority in
favour of the legalisation of same-sex marriage.  Such high turnout
represented a dramatic increase compared to previous
referenda~\cite{MarieVoterTurnout2015}. The enthusiasm of the
electorate was reflected in the activity of on-line social media
platforms, particularly on Twitter which saw a wealth of activity in
the days preceding the referendum~\cite{RTEMarrefLeadingTopic2015}.
Twitter is an online micro-blogging platform where users can post
short messages or {\it tweets} that can be up to 140 characters long;
in Ireland, an estimated $25$\% of adults have a Twitter account, of
which $36\%$ use the service every day~\cite{CraigTwittersWoes2016}.
Users can subscribe to other users' tweets (or {\it follow}); such
following relationships are often asymmetric, if one user follows
another, a reciprocated following relationship does not always
exist~\cite{kwak2010twitterSocNetOrNews}.  In addition to following
each other, there are other ways in which users can publicly interact
such as {\it re-tweeting} (passing forward another user's tweet), and
mentioning each other in tweets.  Twitter has been a popular venue for
the dissemination of information, memes, opinions, and has facilitated
public debate about a variety of subjects~\cite{Gleeson2014,
  Beguerisse2014, Beguerisse2017, alvarez2015sentiment,
  tumasjan2010predicting, metaxas2011not,
  kwak2010twitterSocNetOrNews}. As a result, Twitter has received
considerable attention from researchers who wish to gain insights into
the relationships and mechanisms that govern these social
interactions~\cite{TahaBisedReview2016}.

The use of sentiment analysis to infer the disposition of individuals
or groups towards specific topics is a growing area of interest in
computational social science~\cite{Mislove2010, liu2012survey,
  PangSentimentOpinion2008,TahaBisedReview2016,Dodds2015,
  Bliss2012}. For example, sentiment analysis on Twitter data has been
used to study stock market
fluctuations~\cite{bollen2011twitter,zheludev2015can}, film box-office
performance~\cite{HubermanPredFutureSM2010} and
reviews~\cite{turney2002ThumbsUpOrThumbsDown}, tracking the spread of
influenza~\cite{CristianiniTrackFluSM2010}, and (albeit
controversially) predicting elections~\cite{tumasjan2010predicting,
  OConnor2010tweets, livne2011party, bermingham2011using,
  unankard2014predicting}. Although some of these studies have
well-noted shortcomings~\cite{gayo2012wanted, Gayo-Avello2012}, the
idea of using the content of tweets to gain insight into social
phenomena remains a promising and compelling one. Recent studies,
using carefully constructed methodologies, have successfully leveraged
sentiment to uncover insights into its effect on the spreading of
cascades on Twitter~\cite{alvarez2015sentiment}, and how top
broadcasters send messages with positive sentiment more often than
negative~\cite{Charlton2016InTheMood}.

In this work we combine analyses of sentiment and social structure to
explore Twitter conversations about the Irish marriage referendum. In
particular, we address the following questions:
\begin{itemize}
\item How did users interact with each other on Twitter in the context
  of the Irish Marriage Referendum?
  \item Can user interactions and the sentiment of their tweets help
    us find supporters of voting \textit{yes} (in favour of the
    legalisation of same-sex marriage) and \textit{no} (against it)?
\end{itemize}
To answer these questions, we analyse an extensive dataset of tweets
about the referendum, and the interactions among the users who posted
the tweets (Sec.~\ref{sec:datacollection}). We extract a sentiment
score for each tweet (Sec.~\ref{sec:sentimentAnalysis}), and
incorporate it into the structure of the mention and follower networks
of users (Sec.~\ref{sec:networkConstruction}). These networks enable
the analysis of how the sentiment of users is correlated, and the
proclivity of users with positive/negative sentiment to cluster
together (Sec.~\ref{sec:testingRel}). We use community detection to
partition the users in the mention and follower networks into groups
who communicate more or are generally more interested in each other's
content. We examine these communities from the vantage point of
sentiment analysis to find a parsimonious three-group partition of the
users (Sec.~\ref{sec:communityDetection}). These three groups are
broadly composed of {\it yes} and {\it no} supporters with varying
levels of activity, and starkly different patterns of interaction with
each other (Sec.~\ref{sec:classAccuracy}).  Finally, in
Sec.~\ref{sec:conclusions} we discuss our results and explore
potential future research directions.

\section{Data}\label{sec:datacollection}

\begin{figure}[tp]
  \centering
  \includegraphics[width=0.85\textwidth]{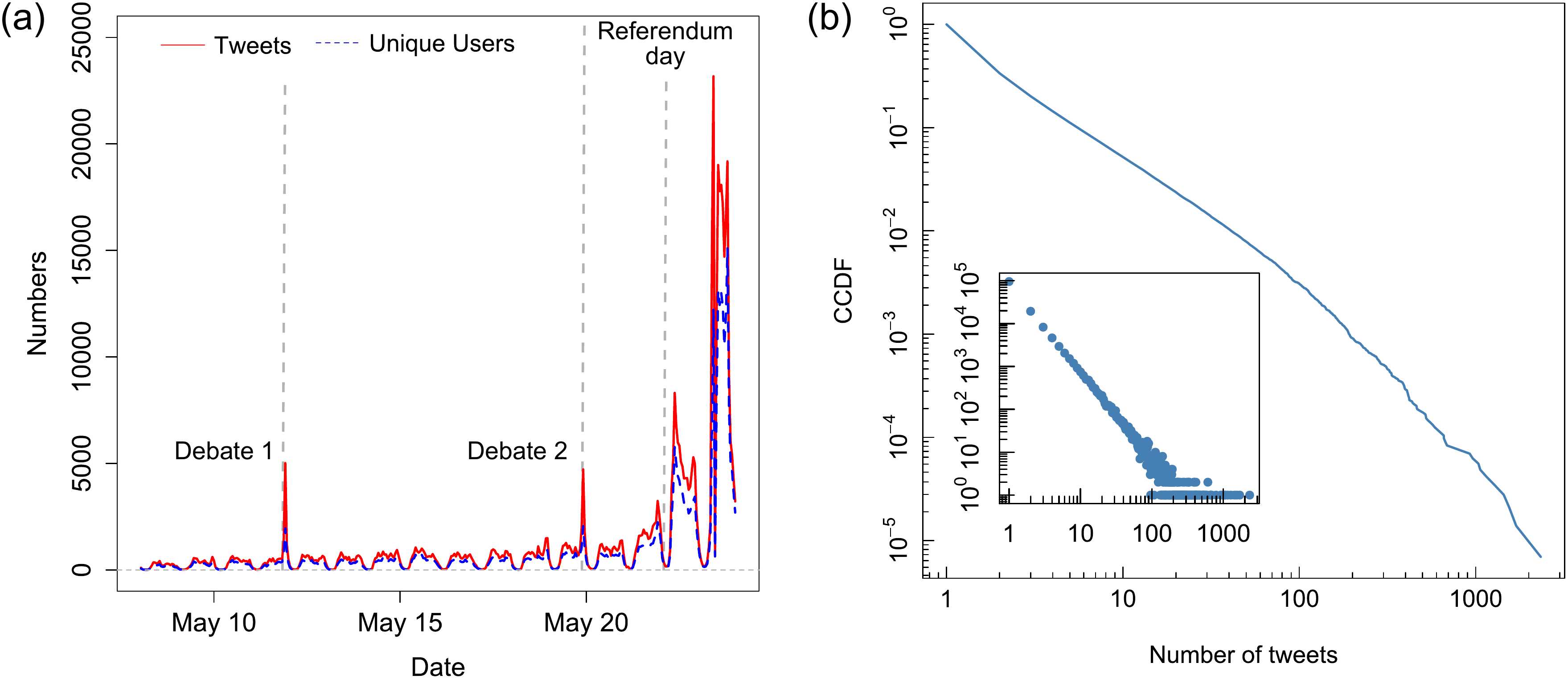}
  \caption{(a) Number of tweets containing the tracked hashtags (solid
    red line) and number of unique users (dashed blue line) in $15$
    minute bins. The volume of tweets increases over time with the
    notable spikes for the two televised debates and the referendum
    day. (b) Complementary Cumulative Distribution Function (CCDF) for
    number of tweets per user on a log-log scale (Inset: the
    Probability Distribution Function (PDF) of the same
    data).}\label{fig:tweetDist}
\end{figure}

The dataset we use in this work consists of {\it every} tweet
containing the hashtags {\tt \#marref} and {\tt \#marriageref} from
May 8 to May 23 $2015$ (one day after the referendum). In total we
collected 499,642 tweets posted by 144,007 unique users (see
Fig.~\ref{fig:tweetDist}(a)). A total of 204,626 tweets were posted
before the referendum day; 88,320 on the day; and 206,696 after. The
peaks observed in Fig.~\ref{fig:tweetDist}(a) coincide with the first
and second televised debates (held on May $11$ and May $19$) and the
referendum day (May $22$), the tallying and announcement of the
results, and subsequent global reaction. Figure~\ref{fig:tweetDist}(b)
shows that the number of tweets per user has a heavy tailed
distribution. The vast majority of users only posted a small number of
tweets with the tracked hashtags, while a small number of users are
responsible for a large volume of tweets. Of the total number of
tweets, 135,370 ($27\%$) were {\it original}, 24,397 ($5\%$) were {\it
  replies} and 339,875 ($68\%$) were {\it retweets}. Broadly speaking,
original tweets are messages that are not in response to another
previously posted tweet (i.e., the content is `new'), replies are
tweets that are posted in response to an existing original tweet, and
retweets are tweets written by others that a user passes along to
his/her followers. In addition users can {\it mention} each other in
their tweets by inserting a user's screen name (technically, replies
and retweets can be seen as specific types of mention tweets).  In our
data there are 388,161 mention tweets ($78\%$ of all tweets), of which
25,732 are original, 23,131 are replies, and 339,298 are retweets. In
addition to tweets, we also obtained the follower relationships of all
users who tweeted one of the hashtags in our data (i.e., a list of
everyone who is followed by the authors of the tweets in our data,
regardless of whether they used the tracked hashtags). These
correspond to 117,669,550 follower links. We also collected user
information such as self-defined location, self-description and how
long the user has been a member of Twitter.

All data was collected by Sinnia using Twitter Gnip Power-Track
API\footnote{\url{https://gnip.com/realtime/powertrack/}} which
returns a complete dataset, not just a
sample~\cite{goel2015StrVirOnlineDiff}. Using the Twitter stream API
has the limitation that as the popularity of a search term (e.g., a
hashtag) increases, the representativeness of the sample
decreases~\cite{Morstatter2013SampleGoodEnough}. By extracting {\it
  all} tweets with the two hashtags, and all user following
relationships we are able to circumvent such sampling issues. There
could be, however, other issues with the data.  For example, it is
possible that our data gathering could miss important tweets or
individuals if they never tweeted using one of the tracked
hashtags. However, due to the ubiquity of the hashtags {\tt \#marref}
and {\tt \#marriageref} in the weeks leading up to the referendum we
are confident that our data is an adequate representation of the
Twitter discourse about the topic.

\section{Sentiment of tweets}\label{sec:sentimentAnalysis}

To quantify how positive or negative a tweet is, we need to compute
the tweet's sentiment score. We do not consider sentiment with the
categorical {\it positive} or {\it negative} labels; instead we
consider sentiment to be a number whose magnitude denotes how positive
or negative the language expressed is~\cite{pang2005SeeingStars}. For
this task we use the open source sentiment algorithm
\textit{SentiStrength}, a lexicon-based sentiment algorithm that
searches for words that have an associated positive or negative
score~\cite{Thelwall2013SentiStrHeart}. SentiStrength provides a score
of both the positive {\it and} the negative emotional charge of a
string of text (in this case, of each tweet in our data).  Positive
scores range from $1$ to $5$, and negative scores from $-1$ to $-5$. A
score of $1$ (or $-1$) indicates that the tweet has no positive (or
negative sentiment), while a score of $5$ (or $-5$) means that the
tweet has the most positive (negative) score possible. See
Appendix~\ref{app:SentiExample} for more details on how sentiment
scores are obtained with SentiStrength.

\begin{figure}[tp]
	\centering
        \includegraphics[width=0.85\textwidth]{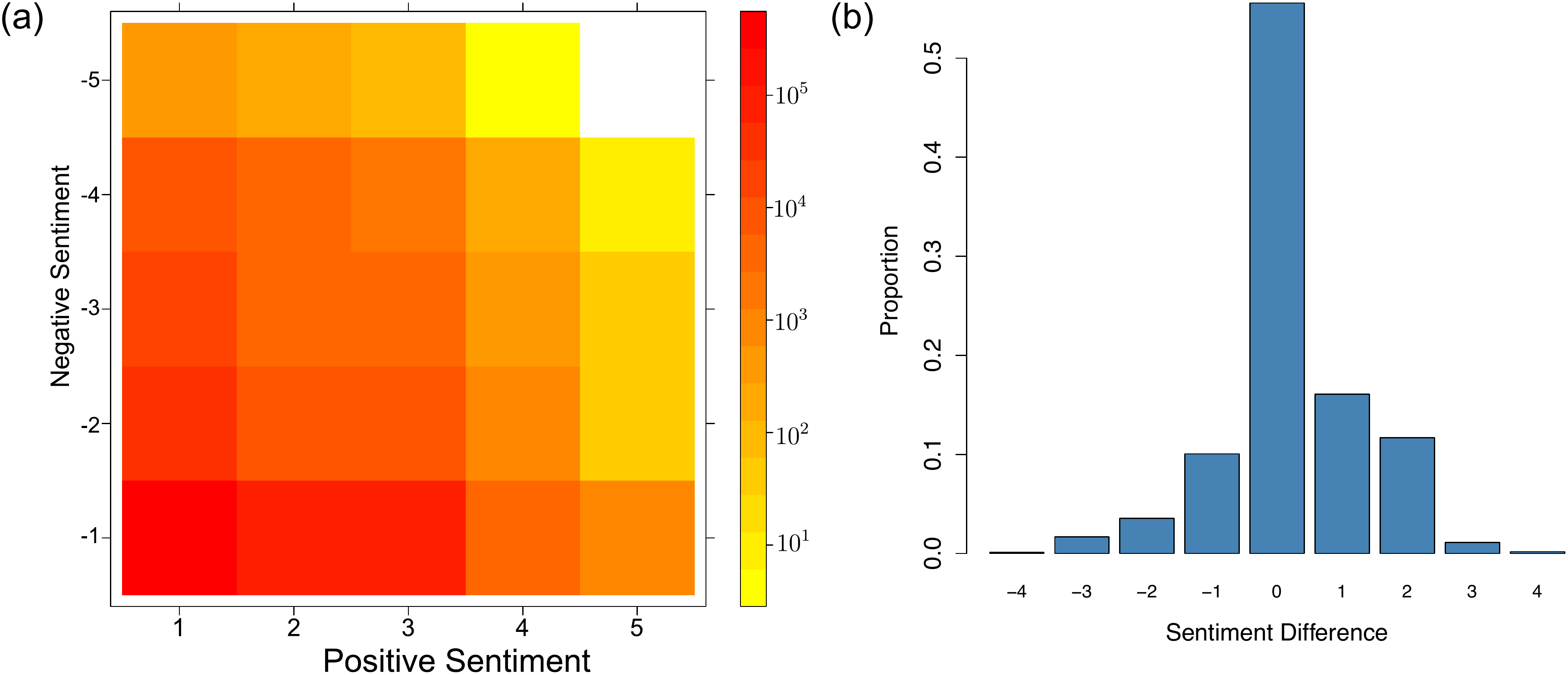}
        \caption{(a) Density plot of the two-dimensional sentiment
          scores of all tweets. (b) Histogram of the difference between
          the positive and negative score of each
          tweet.}\label{fig:sen_diff_scores}
\end{figure}

Figure~\ref{fig:sen_diff_scores}(a) shows the two dimensional
distribution of sentiment scores of all the tweets in our dataset. To
simplify calculations we compute the difference between the positive
and negative scores of each tweet to obtain a one dimensional score
between $-4$ and $4$. A negative score indicates that the tweet
contains stronger negative language than positive, and vice versa for
positive scores. Figure~\ref{fig:sen_diff_scores}(b) shows the
distribution of the unidimensional sentiment scores of all tweets in
the dataset.  About half of all tweets ($55\%$) have a score of zero;
of these the vast majority ($95\%$) have a score of $1$ and $-1$ for
positive and negative language, respectively (i.e., no detected
sentiment), and the rest have balanced positive and negative sentiment
scores. The distribution is roughly symmetric about zero with a slight
positive skew; this observation is consistent with previous reports of
sentiment bias in language~\cite{Dodds2015} and
tweets~\cite{Charlton2016InTheMood}. As noted in the
Appendix~\ref{app:SentiExample}, the SentiStrength scores of a single
tweet can be unreliable, so a single tweet does not provide definitive
information about the user's sentiment. To obtain a more robust
indication of users' sentiment, we can aggregate the scores of all the
tweets produced by one user to obtain a single score. Although
aggregate scores can help overcome some issues, computing a single
score per author neglects the fact that Twitter users often interact
with multiple people, and that the sentiment of these interactions may
vary substantially depending on the counterpart and the nature of the
exchange. Therefore using exclusively a single score per user can lead
to information loss, and provide a misleading indication about the
user's sentiment. To avoid these problems it is necessary to
incorporate the users' interactions into our analysis.

\section{Sentiment aggregation and social structure}
\label{sec:networkConstruction}

Although SentiStrength has been reported to preform well on Twitter
datasets~\cite{Thelwall2013SentiStrHeart}, the nuances and
complexities of human language (for example sarcasm, idioms, negation,
double negatives, and a cavalier attitude towards grammar) make the process
of automatically extracting sentiment a challenging task. In addition,
Twitter users do not exist in isolation, they interact with each other
through mentions, replies and friend/follower relationships. For this
reason it is necessary to incorporate social structure to obtain a
more robust description of the user's disposition with regards to the
marriage referendum. We focus our analysis on two types of Twitter
networks:
\begin{enumerate}
\item Reciprocal mention network: In this network connections exist
  between users who have mentioned each other in tweets containing the
  tracked hashtags.
\item Reciprocal follower network: Connections exist between users who
  {\it follow} each other on Twitter.
\end{enumerate}
The information contained in these networks reflects complementary
aspects of the interactions between users: the reciprocated mention
network includes interactions that arise {\it specifically} from
conversations about the Irish marriage referendum, and are constrained
to the observation period ($8$ to $23$ of May). We are interested in
studying reciprocal mentions because they are a sign of genuine
interactions between
users~\cite{Charlton2016InTheMood,Grindrod150526}. In contrast, the
follower network is not constrained to discussions about the marriage
referendum, nor to the observation period; this network provides a
broader view of how users are interested in each
other. Table~\ref{tab:NetSummary} provides a summary of statistics for
both networks.

\subsection{Construction of the networks}

We construct the directed mention network by searching each users'
tweets for mentions of other users (indicated by a prefixed `@'). A
mention often indicates that the author wishes to draw the attention
of another user to the content of the tweet; this could be original
content directed at a user, a retweet, or a reply.  The announcement
of the referendum results received widespread international attention,
which translated into a large number of tweets from users outside of
Ireland (see Fig.\ref{fig:tweetDist}(a)). We are specifically
interested in detecting {\it yes} and {\it no} supporters, which is
why we further refine our networks to only include tweets generated
before the day of the referendum.  Each mention creates a directed
connection from the author of the tweet to the user who has been
mentioned. We can incorporate sentiment into this network by setting
the weight of the connection to be the sentiment score of the
tweet. When there are multiple directed mentions, we average their
sentiment scores. The resulting network is directed, weighted and
signed (negative weights indicate when the mentions have a
predominantly negative sentiment); it contains $40,812$ unique users,
and $227,203$ directed connections. Note that the users who appear in
this network may not have used one of the tracked hashtags, they only
need to have been mentioned in a tweet containing one of the
hashtags. The average combined in and out degree is $11$, with a
transitivity coefficient of $0.02$. The {\it reciprocal mentions
  network} is the subnetwork in which connected individuals have
mentioned each other in their tweets. This network has $23,713$ edges
($\sim 10\%$ of the mentions in the full network), and $2,830$ users
with non-zero in- and out-degree.

To construct the follower network we obtain the following
relationships between users who authored the tweets in our
dataset. This network has $36,674$ users with $3,309,687$ connections,
of which $1,398,236$ ($42\%$) are reciprocal. The average combined in-
and out-degree is $180$ and the transitivity coefficient is
$0.09$. The full follower network has a different size to the full
mention network because the latter networks starting point was the
users who have authored at least one of the tweets in our database.
Of the $2,830$ users in the reciprocal mention network $2,056$ are in
the largest connected component, of these users $2,047$ users are in
the largest connected component of the follower network.  The final
mention and follower networks contain the users in this $2,047$ node
set with $69,022$ and $173,137$ connections,
respectively. Table~\ref{tab:NetSummary} contains the summary
statistics for the networks.
\begin{table}[tp]
\centering
\begin{tabular}{r|rr|rr}
  \multicolumn{1}{l|}{} & \multicolumn{2}{c|}{Mention} &
  \multicolumn{2}{c}{Follower} \\
  & Full        & Reciprocal    & Full          & Reciprocal    \\ \hline
  Nodes         & 40,812        & 2,047         & 36,674        & 2,047 \\
  Links         & 227,203       & 69,022        & 3,309,687     & 173,137 \\
  Reciprocal links & 23,713     & 22,218        & 1,398,236     & 85,986 \\
  Avg. out degree  & 9          & 34            & 90            & 85     \\
  Transitivity     & 0.02       & 0.13          & 0.09          & 0.28   \\ 
\end{tabular}
\caption{Summary statistics for the mention and follower
  networks.}\label{tab:NetSummary}
\end{table}

To incorporate the sentiment of tweets with the social structure of
the networks described above, we compute four user attributes: the
average in- and out-sentiment ($S_I$ and $S_O$) of each user in the
mentions network, as well as the average in- and out-sentiment of each
user's neighbours ($S^{n}_I$ and $S^{n}_O$). These quantities allow us
to aggregate sentiment scores whilst preserving the heterogeneity of
the user's interactions (e.g., supportive or adversarial discussions).

\begin{figure}[p]
  \centering
   \includegraphics[width=0.85\textwidth]{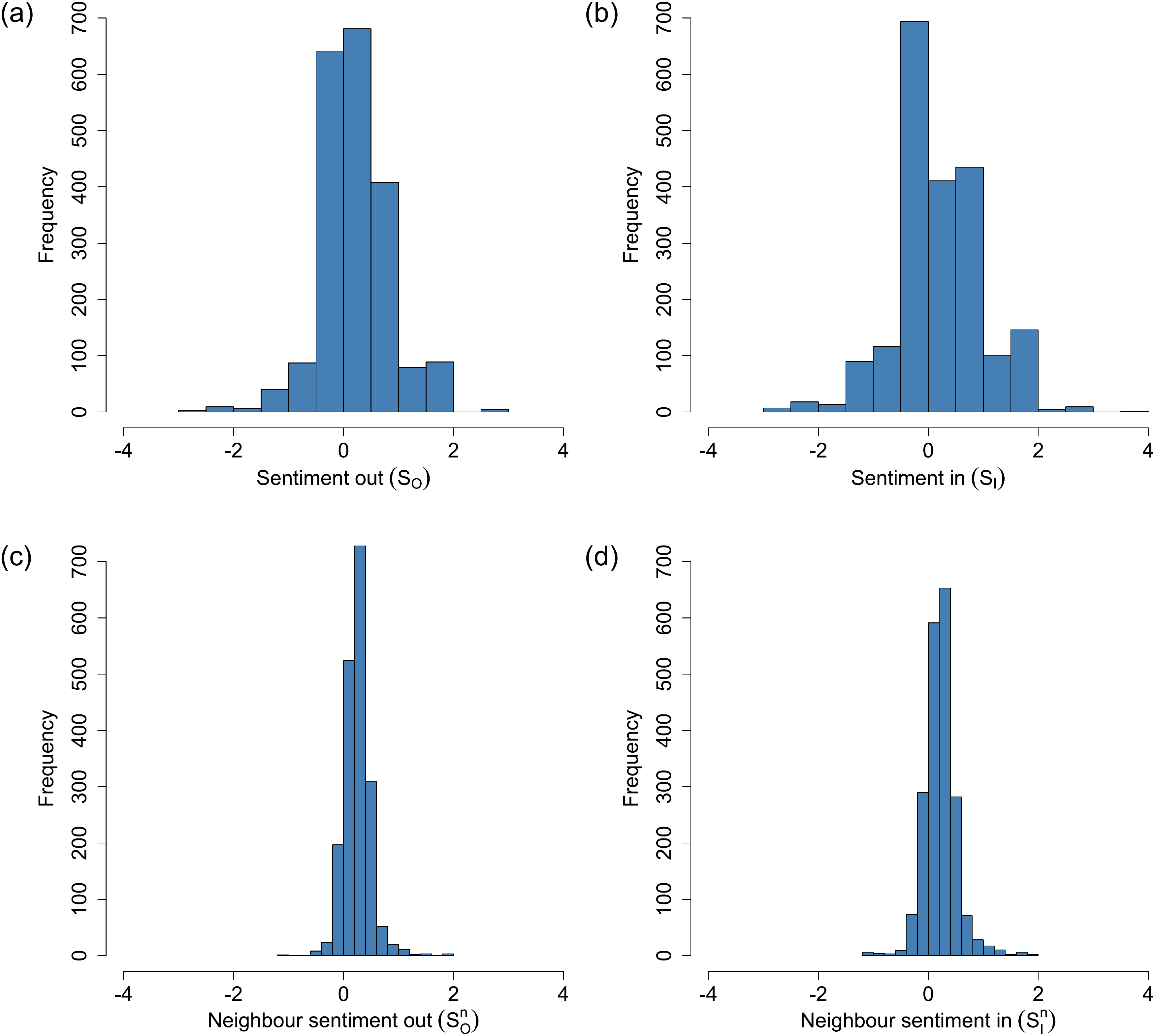}
  \caption{Distribution of the average of users' (a) in-sentiment, (b)
    out-sentiment, (c) neighbour's in-sentiment, and (d) neighbour's
    out-sentiment.}\label{fig:sen_inout}
\end{figure}

Figure~\ref{fig:sen_inout} shows that users tend to have approximately
the same average in- and out-sentiment. The average out-neighbour
sentiment is marginally higher than the average in-neighbour sentiment
($0.26$ and $0.22$, respectively). The average sentiment scores are
centred about zero with a distribution that is approximately
symmetric.

\section{User sentiment alignment}
\label{sec:testingRel}

As discussed in Sec.~\ref{sec:sentimentAnalysis} (and
Appendix~\ref{app:SentiExample}) the sentiment score of a single tweet
is not itself is not entirely reliable; however, the scores of a large
number of tweets can provide a more robust indication of the sentiment
of the corpus. We test this intuition against the null hypothesis that
the sentiment is generated by an inherently random process. 
For example, if the content of the tweets is completely
unrelated to sentiment, or if the sentiment extraction process gives
noisy scores that do not contain any information about the actual
sentiment of the tweets.

More precisely, we would like to determine 1) whether user in- and
out-sentiment scores are correlated; and 2) if users whose tweets have
similar sentiment tend to be clustered in the network. If the
sentiment of the mention tweets that a user sends and receives is
correlated, and users tend to cluster together with others with
similar sentiment, we could then consider sentiment alignment as a
proxy for homophily among users. We can reasonably expect this because
users with a similar {disposition} towards the referendum may
communicate using similar language.  For instance, {\it yes}
campaigners may use phrases that are more positively charged (e.g.,
``vote yes'') more often in their tweets, which results in a higher
positive user sentiment (and vice versa for {\it no} campaigners).

\begin{figure}[tp]
  \centering
  \includegraphics[width=0.45\textwidth]
                  {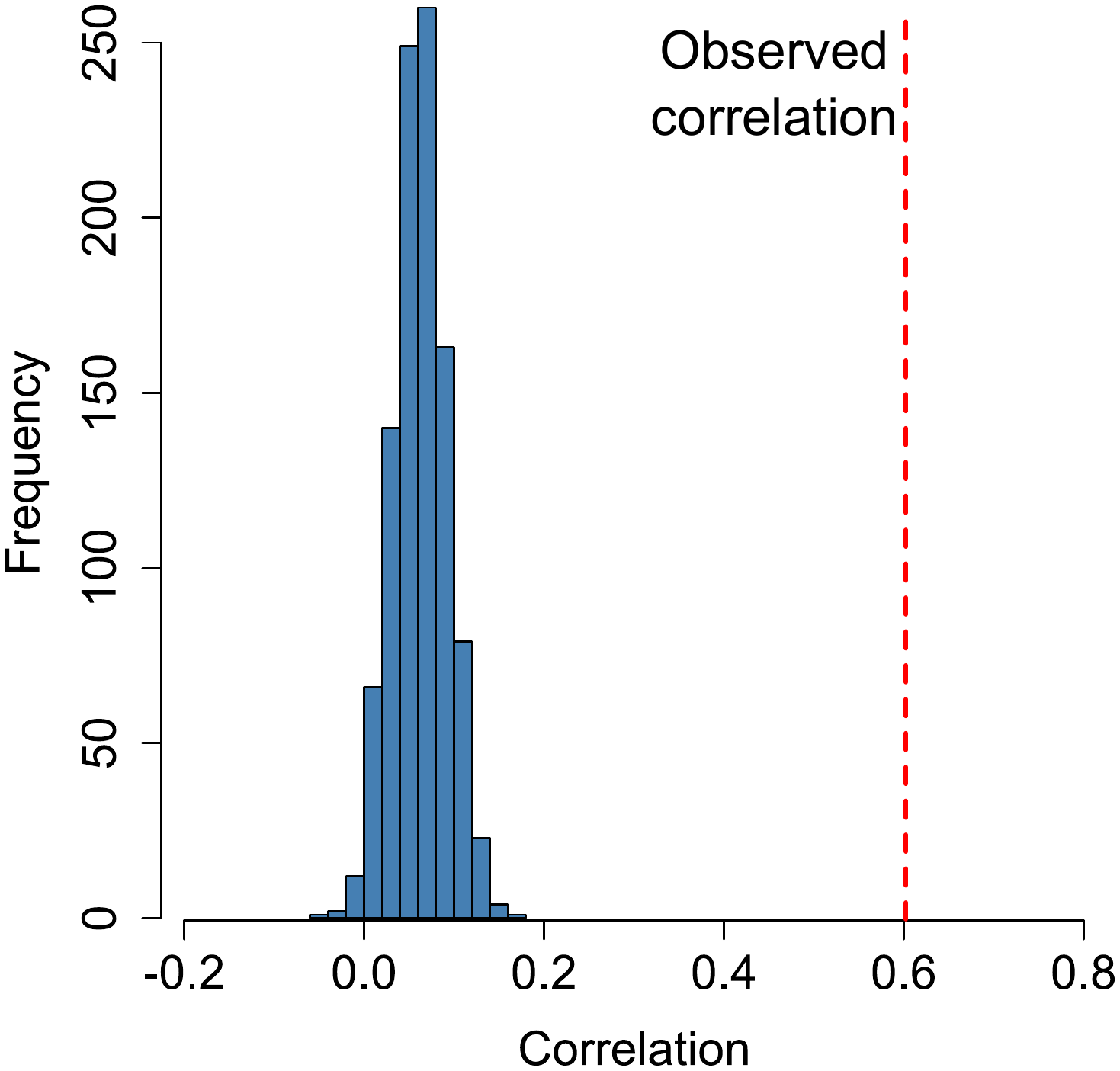}
                  \caption{Distribution of the correlation between
                    $S^r_I$ and $S^r_O$ after 1,000 randomisations
                    (blue bars), and the observed correlation between
                    $S_I$ and $S_O$ in the data (red dashed line).}
  \label{fig:random_cor}
\end{figure}

To answer 1), we examine whether there is a correlation between a
user's in- and out-sentiment.  The Pearson correlation between $S_I$
and $S_O$ is $0.60$, which indicates a moderate linear relationship
between these two nodal attributes~\cite{mukaka2012guide}. To confirm
that this correlation is not due to chance alone, we use a procedure
based on redistributing the sentiment of a user's tweets. The
randomisation procedure is: 
\begin{itemize}
\item Sample a sentiment score for each connection from the observed
  distribution of link scores with replacement. This keeps the network topology intact.
\item Calculate the average randomised in- and out-sentiment of each user ($S^r_I$ and
  $S^r_O$). 
\item Calculate the correlation coefficient between $S^r_I$ and
  $S^r_O$ in the resampled network.  
\end{itemize}
Figure~\ref{fig:random_cor} shows the comparison of the resulting
distribution of the correlation between ($S^r_I$ and $S^r_O$) after
1,000 iterations of the procedure with the observed correlation of $S_I$ and $S_O$ in
our data.  This result indicates that there is a nontrivial
correlation between the sentiment of what a user tweets and receives.

\begin{figure}[tp]
	\centering
	\includegraphics[width = 0.7\textwidth]
                        {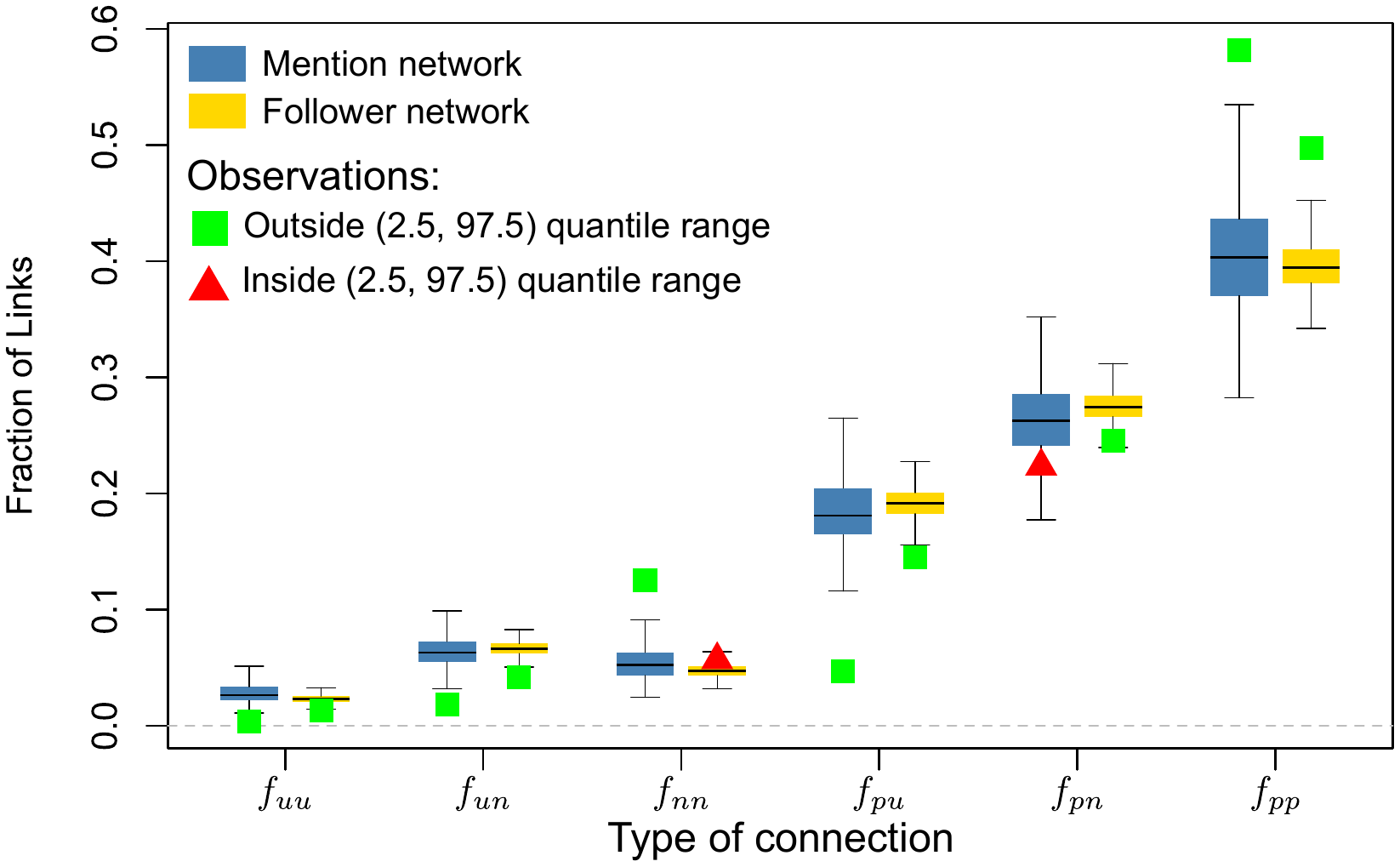}
                        \caption{Result of the randomisation tests in
                          the the mention (blue box plots) network and
                          the follower network (yellow box plots). The
                          green squares and red triangles mark the
                          observed fraction of links in the
                          data. Green squares indicate that the
                          observed fraction of connections falls
                          outside the lower $2.5\%$ and upper $97.5\%$
                          quantiles of the randomised distribution
                          (i.e., it is unlikely to arise by chance
                          alone); red squares indicate that the
                          observed fraction falls inside the lower
                          $2.5\%$ and upper $97.5\%$ quantiles of the
                          randomised
                          distribution.}\label{fig:randomisationTests}
\end{figure}

To answer 2) we investigate whether users with similar sentiment are
clustered together in the mention and follower networks. The observed
correlation between $S_I$ and $S_O$ suggests that users may be more
likely to be connected to other users with similar sentiment scores.
We create three coarse class labels for users according to their
sentiment --- aggregate scores above zero are ``positive'', scores
less than zero are ``negative'' and scores equal to zero are
``unknown'' --- and we find the fraction of links connecting users of
these broad sentiment labels.  We denote the fraction of links between
positive and positive users as $f_{pp}$, the fraction of links between
positive and negative users as $f_{pn}$, between positive and unknown
users as $f_{pu}$, and so on.  In total there are six types of links:
$f_{pp}$, $f_{pn}$, $f_{pu}$, $f_{nn}$, $f_{un}$ and $f_{uu}$. We
randomise the class labels of each user by sampling from the observed
distributions with replacement, and recalculate the fraction of
connections; we repeat this process $1,000$ times. As before, we
compare the randomised distributions of the fractions with the
observed fraction in our data; Figure~\ref{fig:randomisationTests}(a)
shows the results for this procedure.

The randomisation test in the mention network (blue box plots in
Fig.~\ref{fig:randomisationTests}) shows that it is highly unlikely
that the $f_{pp}$, $f_{pu}$, $f_{nn}$, $f_{un}$ and $f_{uu}$ observed
in the mention network arise from chance. There are fewer connections
involving unknown users ($f_{un}$ and $f_{uu}$ and $f_{pu}$) than we
would expect by chance.  In contrast, the connections between positive
users ($f_{pp}$) and negative users ($f_{nn}$) are higher than
expected.  The faction of connections between positive and negative
users ($f_{pn}$) is less than what we would expect (below the $25\%$
quantile), although this result is less robust than the rest.  This
analysis shows that users tend to mention in their tweets other users
with similar sentiment more frequently than we would expect by chance.
The same analysis in the follower network (yellow box plots in
Fig.~\ref{fig:randomisationTests}) paints a broadly consistent
picture.  We find more links between positive users, fewer links
between positive and negative, and fewer links involving unknown users
than we would expect by random chance.

This analysis provides evidence of a relationship between users' $S_I$
and $S_O$, and their preference to engage with users of a similar
sentiment, and supports the notion that in this case sentiment can be
a proxy for homophily. Labelling users as positive, negative or
unknown, according to the sign of their $S_O$ may seem too coarse; to
test this we performed a similar analysis in which we split the users
according to their quantile, or by above/below the mean or median (see
Appendix~\ref{app:randomAlternative}). These tests produced the same
results we present in this section.

\section{Communities and sentiment}\label{sec:communityDetection}

\begin{figure}[tp]
  \centering
  \includegraphics[width=0.95\textwidth]{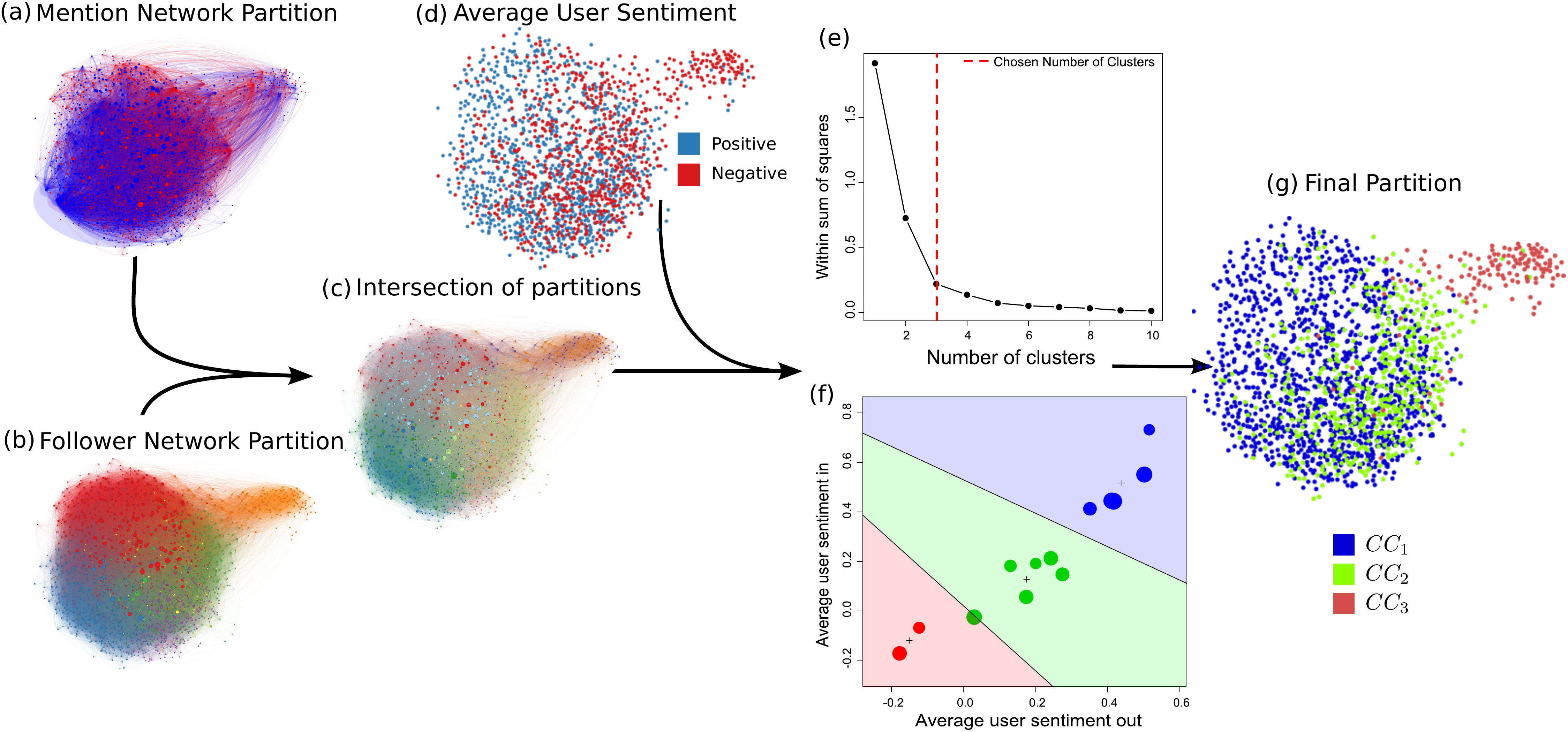}
  \caption{Schematic of our analysis of communities and
    sentiment.  Communities in the mention network (a) and follower
    (b) networks.  The intersection of the communities in both
    networks is shown in (c).  Mention network with nodes coloured
    according to sentiment (d).  The $k$-means clustering of the
    sub-communities according to their sentiment reveals three broad
    clusters (e).  The relationship between in- and out-sentiment of
    each sub-community and cluster membership is shown in (f). The
    size of each marker is proportional to the size of each sub
    community; crosses indicate the centroid of each cluster.  Final
    partition of users into three ``community clusters'' in the
    mention network.}\label{fig:communityAma}
\end{figure}

We are interested in finding groups of users that are not only tightly
connected in both networks, but also whose tweets have similar
sentiment. For this task, we extract the communities in each network,
and enrich it with the analysis of connection patterns from the
previous section.  We use Markov
Stability~\cite{Delvenne2010,delvenne2013stability} to obtain a robust
partition of the mentions network into $17$ communities, and a
partition of the follower network into $7$ communities
(Fig.~\ref{fig:communityAma}(a) and (b)). 

The communities in the mention network arise specifically around
conversations between users; the links consist of mention tweets
containing the tracked hashtags posted during the observation
period. In contrast, the communities in the follower network arise
from the declared interest of a user in receiving another users'
posts, which is not necessarily restricted to the context of the
marriage referendum.  In essence, we now seek to find a new grouping
of users based on both partitions, and use the sentiment scores to
construct a measure of similarity.

To accomplish this task, we intersect the partitions of the two
networks to obtain $62$ sub-communities
(Fig.~\ref{fig:communityAma}(c)). Each of these new groups contains
users that are in the same community in both networks; these users are
not only more broadly interested in each other (because they follow
each other), but also had conversations about the referendum.  Then,
we calculate the average in- and out-sentiment and neighbour sentiment
$\bar{S_I}_i$, $\bar{S_O}_i$, $\bar{S^{n}_I}_i$ and $\bar{S^{n}_O}_i$
in each sub-community $i\in{1, \dots , 62}$. As we noted in the
previous section, we consider sentiment as a proxy for homophily
between users; therefore we use aggregate sentiment scores as an
indication of similarity between the $62$ sub-communities. However,
$49$ of these sub-communities have $20$ users or fewer ($204$ users in
total).  Because sentiment scores of individual tweets are a noisy
signal and these communities are small, we are unable to provide a
robust statistical description in these communities.  To limit the
effect of this noise we remove these sub-communities and proceed to
analyse the remaining $13$ sub-communities.  This procedure is
illustrated in Fig.~\ref{fig:communityAma}.

We use $k$-means clustering to group the sub-communities according to
the euclidean distance between the average in- and out-sentiment and
neighbour sentiment scores of each sub-community. To choose the number
of clusters we locate the bend in the plot of the total within sum of
squares of the sentiment difference of the members of the groups
(Fig.~\ref{fig:communityAma}(e)).  A marked flattening of the graph
suggests that a finer clustering is not considerably better at
segregating sub-communities into distinct groups than a more
parsimonious clustering with fewer groups. The appropriate number of
clusters is found at the ``elbow'' of the
graph~\cite{Aggarwal2013Clustering}, which in this case is three.
Figure~\ref{fig:communityAma}(f) shows the three regions in which we
have classified the sub-communities. We call these clusters of
sub-communities {\it community clusters}: $CC_1$ with $1,064$ users,
$CC_2$ with $604$, and $CC_3$ with $155$. The community cluster $CC_1$
has the highest in- and out-sentiment, followed by $CC_2$ and lastly
$CC_3$.

\begin{table}[tp]
\centering
\begin{tabular}{r|rr|rr|rr}
  \multicolumn{1}{c|}{} & \multicolumn{2}{c|}{$CC_1$}
  & \multicolumn{2}{c|}{$CC_2$} & \multicolumn{2}{c}{$CC_3$} \\
& Mention & Follower & Mention & Follower & Mention & Follower  \\ \hline
Users & \multicolumn{2}{c|}{1,064} & \multicolumn{2}{c|}{604}
& \multicolumn{2}{c}{155}    \\
Links  & 32,076  & 85,302  & 22,333  & 45,799  & 8,119  & 6,409  \\
Reciprocal links & 12,855  & 44,890  & 5,527  & 22,171  & 2,582 & 3,163  \\
Avg. out degree  & 30 & 80 & 37 & 76 & 52 & 41  \\
Transitivity   & 0.15 & 0.35 & 0.15 & 0.27 & 0.45  & 0.57        
\end{tabular}
\caption{Summary statistics for each community cluster. Note that the
  number of reciprocal links and transitivity are calculated for each
  community clusters network in isolation.}\label{tab:summaryCC}
\end{table}

\begin{figure}[tp]
  \centering
  \includegraphics[width=0.85\textwidth]{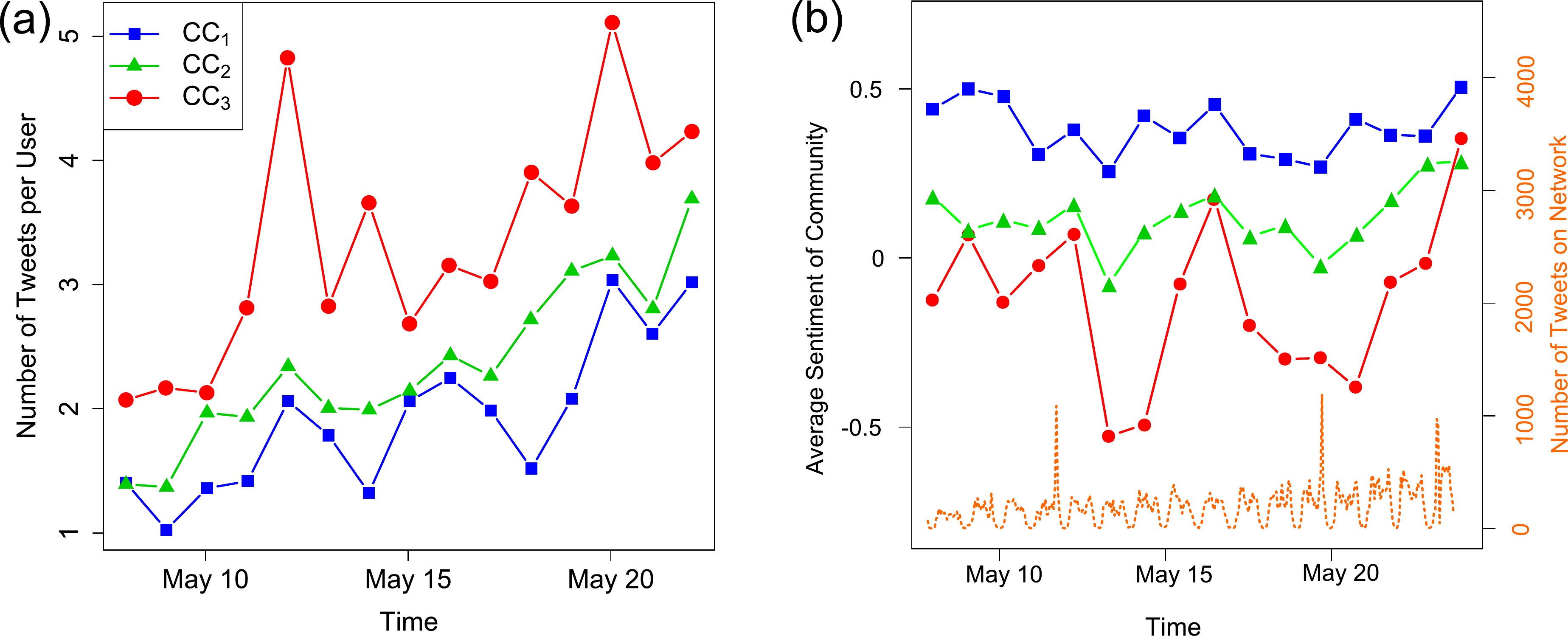}
  \caption{(a) Average number of tweets per user per day in each
    community cluster. (b) Out sentiment of each community cluster
    over time (left y-axis), and number of tweets (right y-axis,
    orange dotted line)}\label{fig:TSAverageTPU}
\end{figure}

Table~\ref{tab:summaryCC} contains the summary statistics for each of
these community clusters: $CC_1$ has the lowest average out degree in
the mention network, followed by $CC_2$, and $CC_3$ has the
highest. The clusters $CC_3$ and $CC_2$ are the most active;
Fig.~\ref{fig:TSAverageTPU}(a) shows that they consistently have the
highest number of tweets per users. Cluster $CC_3$ is the most tightly
connected of the three, with a high transitivity
coefficient in both the mention and follower network. These community
clusters are also consistently stratified by the sentiment of their
tweets over time (Fig.~\ref{fig:TSAverageTPU}(b)).

\section{Support for the {\it yes} and {\it no} sides in the
community clusters}
\label{sec:classAccuracy}

Are the community clusters that we found in the previous section
representative of {\it yes} or {\it no} supporters? To find out we
sample 358 ($20\%$) users at random and classify manually them as
either \textit{yes}, \textit{no} and \textit{unaligned} voters. To
classify each user we examine their Twitter biography
(self-description) and all their tweets in our dataset. If an account
has no obvious leaning, such an automated account (e.g., a {\it bot});
an institutional account, or an impartial journalist, we classify it
as {\it unaligned}. After classifying all the users in our sample, we
examine the composition of each community
cluster. Table~\ref{tab:comBD} shows how the {\it yes}, {\it no} and
{\it unaligned} users are distributed across the sample from each
community cluster. See Appendix~\ref{app:accuracy} for a detailed
outline of the classification procedure.

\begin{table}[tp]
\centering
\begin{tabular}{lr|rrr|r}
          &  & \multicolumn{3}{c|}{Community cluster} & \\
          &  & $CC_1$ & $CC_2$ & $CC_3$ & Total \\ \hline
          & Yes               & 183    & 114    & 6      & 303   \\
 Alignment & No                & 1      & 2      & 23     & 26    \\
          & Unaligned         & 21     & 5      & 3      & 29    \\ \hline
          & Total             & 205    & 121    & 32     & 358  
\end{tabular}
\caption{Number of sampled {\it yes}, {\it no} and {\it unaligned}
  supporters in each community cluster.}\label{tab:comBD}
\end{table}

Users that support the {\it yes} side are predominantly found in
community clusters $CC_1$ and $CC_2$ ($89\%$ of the users in $CC_1$
and $96\%$ of users in $CC_2$ in the sample). Users that lean towards
the {\it no} side are concentrated  in $CC_3$ ($71\%$ of the
users in $CC_3$ from the sample). The unaligned users are mostly in
$CC_1$ ($10\%$) and $CC_3$ ($9\%$), whereas $CC_2$ has the fewest
($4\%$).  We categorise each community cluster according to the
prevalence of {\it yes} and {\it no} leaning accounts; this achieves
an accuracy of $89\%$, and a balanced
accuracy~\cite{brodersen2010balanced} of $81\%$ (see
Appendix~\ref{app:accuracy}).

\begin{figure}[tp]
  \centering
  \includegraphics[width=0.9\textwidth]
                  {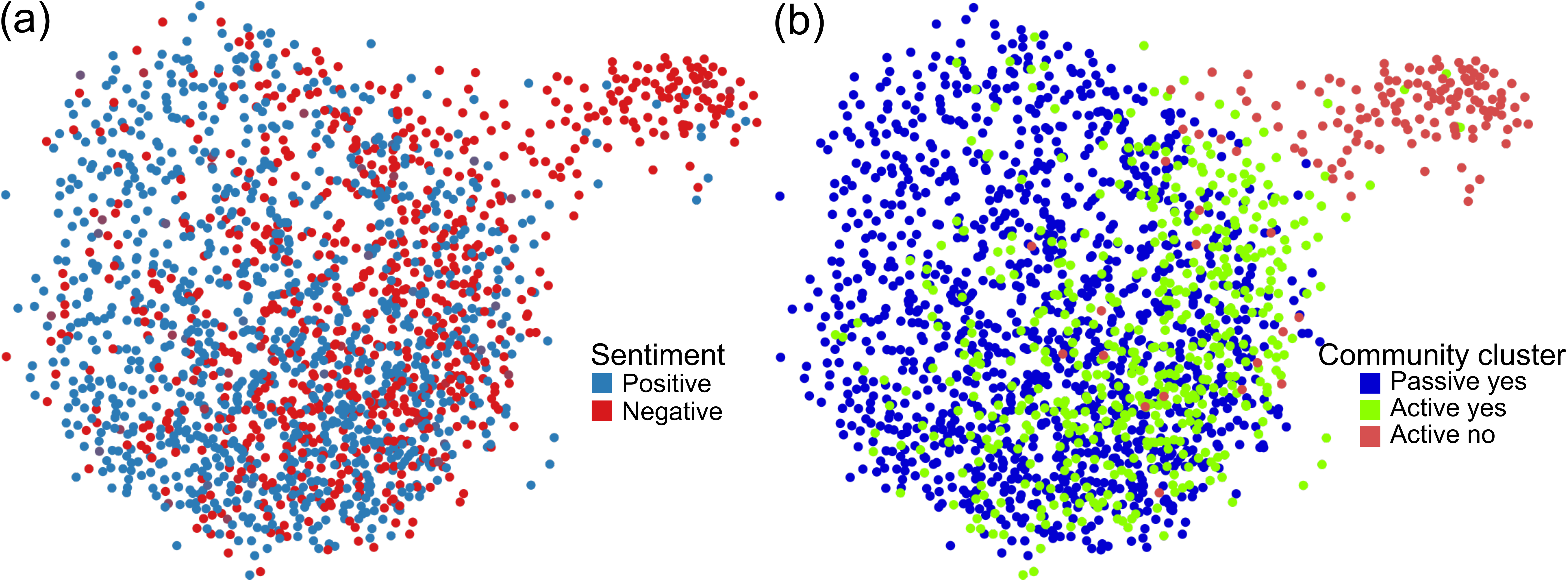}
                  \caption{Reciprocal mention network in which the
                    nodes are coloured by sentiment (a) and the final
                    community-cluster partitions labelled by the side
                    they support in the referendum (b). Edges removed
                    for clarity.}\label{fig:networkViz}
\end{figure}

As we observed in Sec.~\ref{sec:communityDetection}, the community
clusters have varying levels of activity: members of $CC_2$ and $CC_3$
post twice as many mention tweets as $CC_1$ over the observation
period. With these activity levels, in combination with the distribution of
support in Table~\ref{tab:comBD}, we label community cluster
$CC_1$ as {\it Passive Yes}, $CC_2$ as {\it Active Yes}, and $CC_3$ as
{\it Active No}. Figure~\ref{fig:networkViz} shows these
classifications displayed on the layout of the mention network,
alongside their sentiment.

\subsection{Activity of community clusters}
\label{sec:results}

\begin{figure}[tp]
  \includegraphics[width = 0.95\textwidth]
                  {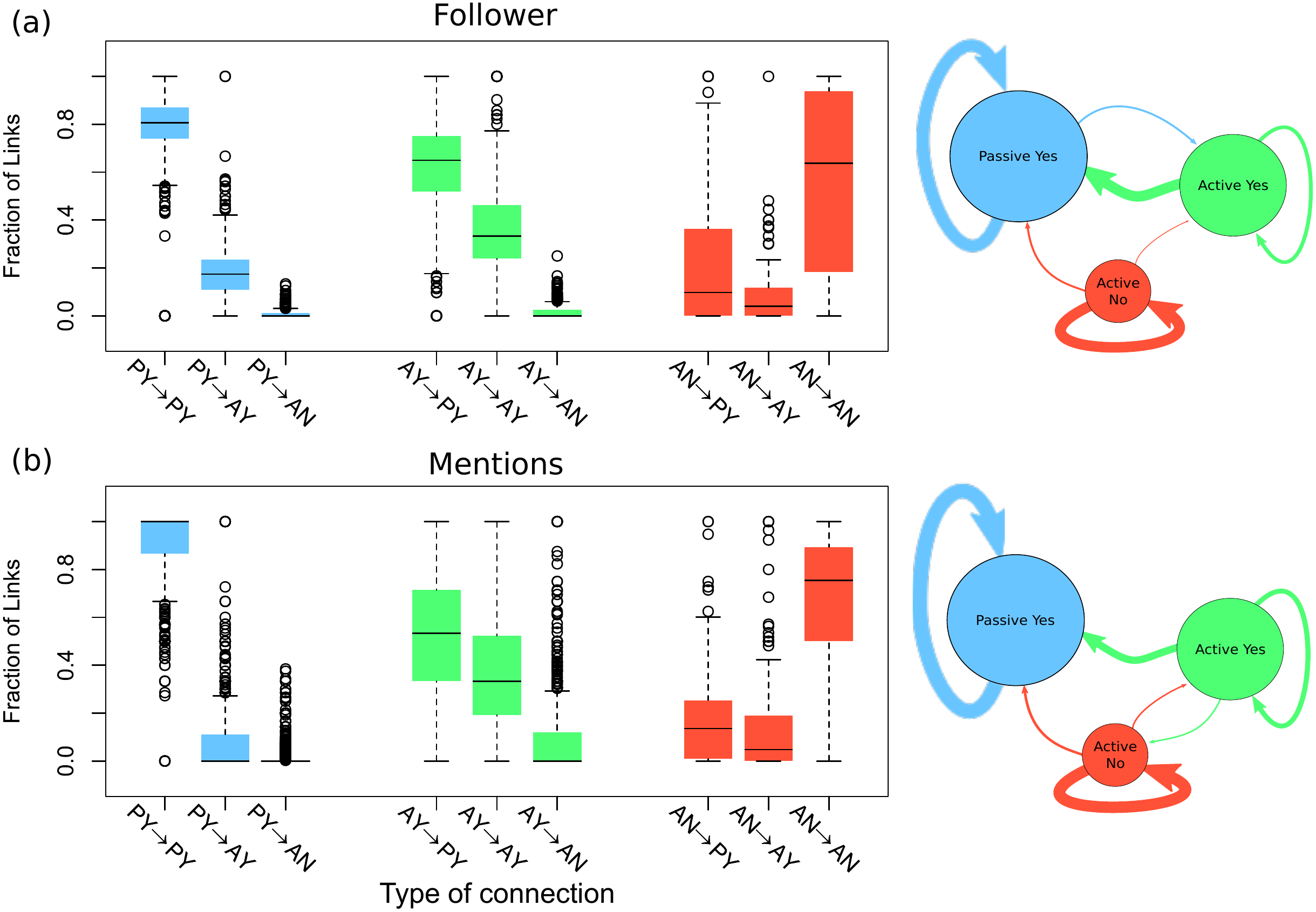}
	          \caption{Fraction of connections between users in
                    the three community clusters for the follower (a)
                    and mention (b) networks.}\label{fig:baxplotsDist}
\end{figure}

We examine which community clusters interact more frequently through
mentions and friend/follower links, the difference in the type of
mention used (original, reply or retweet), and the sentiment of the
interactions between community clusters.
Figure~\ref{fig:baxplotsDist}(a) shows that users in the {\it Passive
  Yes} and {\it Active No} community clusters tend to follow mostly
users within their own group ($80\%$ and $61\%$ of user links,
respectively), whereas users in the {\it Active Yes} cluster follow a
disproportionately large number of users from {\it Passive Yes}
($66\%$). Users in both {\it Yes} communities, on average, follow only
a small number of users in {\it Active No} ($0.7\%$ and $2.6\%$
respectively).  This pattern also appears in the mention network,
where most connections are between members of the same community
cluster (Fig.~\ref{fig:baxplotsDist}(b)). The strongest interaction
between community clusters consists of connections between users in
the {\it Yes} groups in both networks.

\begin{table}[tp]
\centering
\begin{tabular}{ll|rrr|rrr}
   &  & \multicolumn{3}{l|}{Mention tweets} & \multicolumn{3}{l}{Porportion of cluster's tweets} \\
From        & To          & Original     & Reply    & Retweet    & Original          & Reply         & Retweet         \\ \hline
Passive Yes & Passive Yes & 5302         & 1755     & 21740      & 0.16              & 0.06          & 0.68            \\
Passive Yes & Active Yes  & 206          & 306      & 2152       & 0.01              & 0.01          & 0.07            \\
Passive Yes & Active No   & 139          & 168      & 308        & 0.00              & 0.00          & 0.01            \\ \hline
Active Yes  & Passive Yes & 1200         & 1205     & 10130      & 0.05              & 0.05          & 0.45            \\
Active Yes  & Active Yes  & 380          & 1935     & 4648       & 0.02              & 0.09          & 0.21            \\
Active Yes  & Active No   & 286          & 1948     & 601        & 0.01              & 0.09          & 0.03            \\ \hline
Active No   & Passive Yes & 361          & 458      & 753        & 0.04              & 0.06          & 0.09            \\
Active No   & Active Yes  & 47           & 939      & 257        & 0.01              & 0.12          & 0.03            \\
Active No   & Active No   & 310          & 649      & 4345       & 0.04              & 0.08          & 0.54           
\end{tabular}
\caption{Type of communication channel used between community
  clusters. Proportions are given for the total tweets originating
  from each group.}\label{tab:typeCom}
\end{table}

We also examine which type of mentions (original, replies or retweets)
are used by the members of each group in their interactions. All
community clusters retweet more often than they produce original
messages or replies (Table~\ref{tab:typeCom}). Unsurprisingly, retweet
connections occur most often between groups where there is already a
high number of follower connections. We observe a similar situation
with original mention tweets.  Interestingly, reply tweets do not
follow this trend; these messages tend to be sent to community
clusters where there are few follower links to the source cluster.
The users in the {\it Active Yes} and {\it Active No} community
clusters produce the most reply tweets: $24\%$ and $26\%$ of their
tweets are replies, respectively.

Of the total number of tweets sent between the {\it Active Yes} and
{\it Active No} community clusters, $9\%$ and $12\%$ respectively
correspond to replies. This finding is surprising for two reasons:
Firstly, there are very few follower connections between the two
groups, which means that these messages bridged a gap between groups
that do not typically interact.  Secondly, these groups are
ideologically opposed to each other. The {\it Passive Yes} community
cluster, on the other hand, only sent $1.4\%$ of its tweets in the
form of replies to other community clusters. The two active {\it Yes}
and {\it No} community clusters produced $73\%$ of all replies,
although they represent only $35\%$ of all users.

\begin{table}[tp]
\centering
\begin{tabular}{ll|rrr}
From        & To          & Original & Reply & Retweet \\ \hline
Passive Yes & Passive Yes & 0.99     & 0.96  & 0.97    \\
Passive Yes & Active Yes  & 0.99     & 0.96  & 0.97    \\
Passive Yes & Active No   & 0.68     & 0.49  & 0.53    \\ \hline
Active Yes  & Passive Yes & 1.00     & 1.00  & 1.00    \\
Active Yes  & Active Yes  & 1.00     & 0.98  & 1.00    \\
Active Yes  & Active No   & 0.70     & {\bf 0.59}  & 0.76    \\ \hline
Active No   & Passive Yes & 0.96     & 0.81  & 0.75    \\
Active No   & Active Yes  & 0.83     & {\bf 0.47}  & 0.60    \\
Active No   & Active No   & 0.94     & 0.76  & 0.97   
\end{tabular}
\caption{Fraction of mention tweets that occurred between nodes that
  are connected in the follower network.}
\label{tab:FracAlongPipe}
\end{table}

We also calculate the fraction of original, replies and retweets that
occurred in the presence of a follower
link. Table~\ref{tab:FracAlongPipe} shows that, of all the reply
tweets between the active {\it Yes} and {\it No} communities, only
$56\%$ and $52\%$ occurred when there was a follower link between the
users. This is yet another indication that users in these two groups
were more likely to engage with each other, even in the absence of
strong structural ties. These results are consistent with the notion
that although the marriage referendum was a heated topic on Twitter,
the engagement between users with different views was limited to a
small subset of highly active users.  Note that we only study tweets
made using two hashtags; it is possible that the actual
number of replies was higher.

Given the differences in the type of mentions between the community
clusters, we are also interested in knowing whether the sentiment of
the connections varies depending on the source and the target group.
We unfold the average out sentiment of each user ($S_O$) to see the
 scores of tweets directed at each community cluster.
\begin{figure}[tp]
\centering
\includegraphics[width = 0.95\textwidth]
                {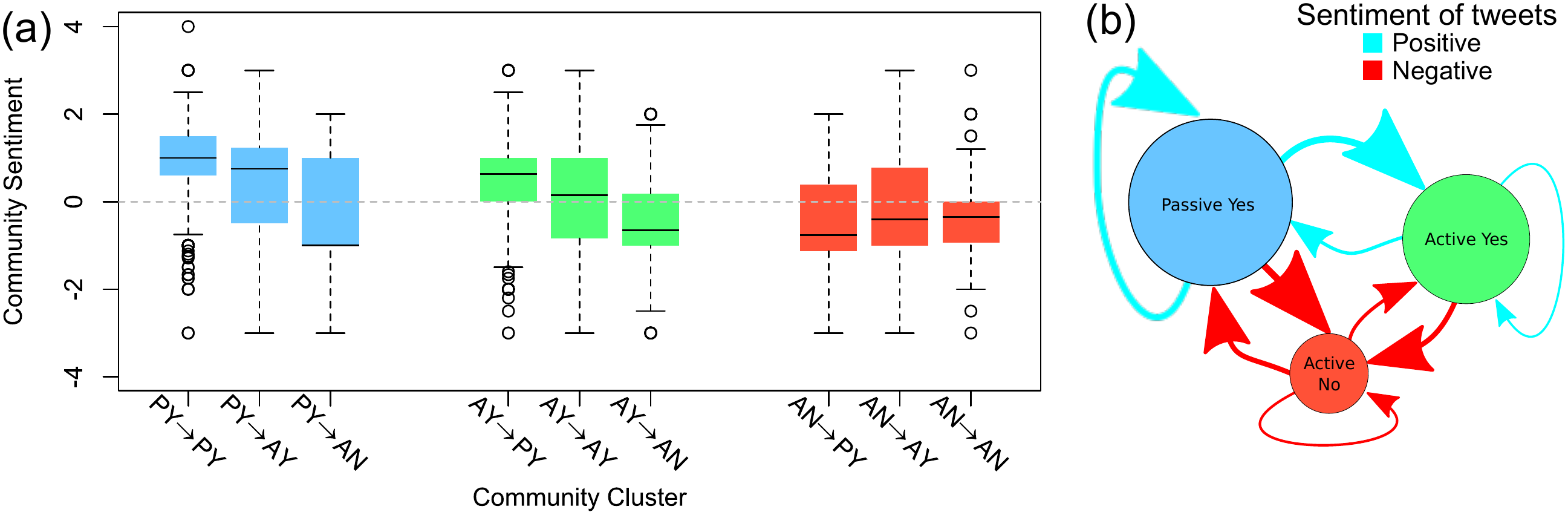}
	        \caption{Box plots with the sentiment of the
                  interactions between community clusters (a). On the
                  network in (b)  we see an illustration of these
                  boxplots in the mention network. The size and colour
                  of the arrow are proportional to the mean sentiment
                  of the connections from each community
                  cluster.}\label{fig:sentiment}
\end{figure}
Figure~\ref{fig:sentiment} shows that interactions with the Active No
community cluster have, on average, more negative sentiment than other
interactions. The interaction between the {\it Active Yes} and {\it
  Active No} clusters is overwhelmingly negative. Over $50\%$ of users
from both active community clusters use language with negative
sentiment in their mention tweets sent between each other. The
interactions of the {\it Passive Yes} cluster with itself, on the
other hand, are overwhelmingly positive; over $50\%$ have a positive
sentiment score. The opposite is true for any interaction of the {\it
  Active Yes} with the {\it Active No} community cluster. This is the
main feature that allows us to distinguish the Active Yes community
cluster from the Passive Yes. The interactions between Active and
Passive Yes are almost all positive, and consist mostly of retweets.
In contrast, the interactions between Active Yes and Active No are
typically negative, and consist of replies.

\section{Conclusion}\label{sec:conclusions}

We have investigated the relationship between sentiment and social
structure in the context of the Twitter discussion about the 2015
Irish Marriage referendum. We computed the sentiment scores of
$204,626$ tweets posted by $36,674$ users, and constructed follower
and mention networks among users in which the weight of the
connections corresponds to the sentiment of the interactions.  Our
results show that although the sentiment score of individual tweets
can be noisy, it can be aggregated successfully using networks to
study the interactions between users in a mention and follower
network.  We performed extensive statistical tests to study the
relationship between the sentiment of users' tweets and their
interactions, both in general (i.e., friend/follower) and
topic-specific (i.e., from tweets about the referendum). The
correlation between the sentiment of mentions that a user sends and
receives (the in- and out-sentiment) is positive and robust to
randomisation tests. Furthermore, our tests show that users in the
mentions network with positive and negative aggregate sentiment scores
are more likely to be connected to users with similar sentiment than
would occur by chance; positive users are also more likely to follow
each other.  The community structure of the networks shows that users
with similar sentiment tend to be clustered together.  We combined
sentiment scores with the networks' communities to find three distinct
groups of users that we classified as either {\it yes} or {\it no}
supporters based on the content of their tweets and sentiment, and as
{\it active} or {\it passive} based on their activity. Interestingly,
many of the mentions between the users in the {\it yes} and {\it no}
groups occurred in the absence of friend/follower links, which
indicates the existence of topical dialogue across ideological
lines. These results show that sentiment and social structure are
related yet distinct, and must be studied together to understand the
disposition of users around topics of interest. This work can be
extended in a number of directions, for example by combining sentiment
analysis with topic modelling and additional user features (such as
demographics, age, gender, or income) to obtain a more accurate
picture of user disposition. We anticipate that this work will also
provide a basis for incorporating sentiment in opinion dynamics models
and the analysis of retweet cascades, and to investigate the
calibration of polling data using social structure.

\section*{Acknowledgments}
This work was partially funded by Science Foundation Ireland (awards
11/PI/1026 and 12/IA/I683). MBD acknowledges support Oxford-Emirates
Data Science Lab and a James S. McDonnell Foundation Postdoctoral
Program in Complexity~Science/Complex~Systems Fellowship Award
(\#220020349-CS/PD Fellow). The authors thank Kevin Burke and Pete
Grindrod for advice and fruitful discussions.


\appendix

\setcounter{figure}{0}
\makeatletter 
\renewcommand{\thefigure}{S\@arabic\c@figure}
\makeatother

\setcounter{table}{0}
\makeatletter 
\renewcommand{\thetable}{S\@arabic\c@table}
\makeatother

\section{Sentiment extraction with SentiStrength}
\label{app:SentiExample}

\begin{figure}[tp]
  \centerline{\includegraphics[width=0.85\textwidth]{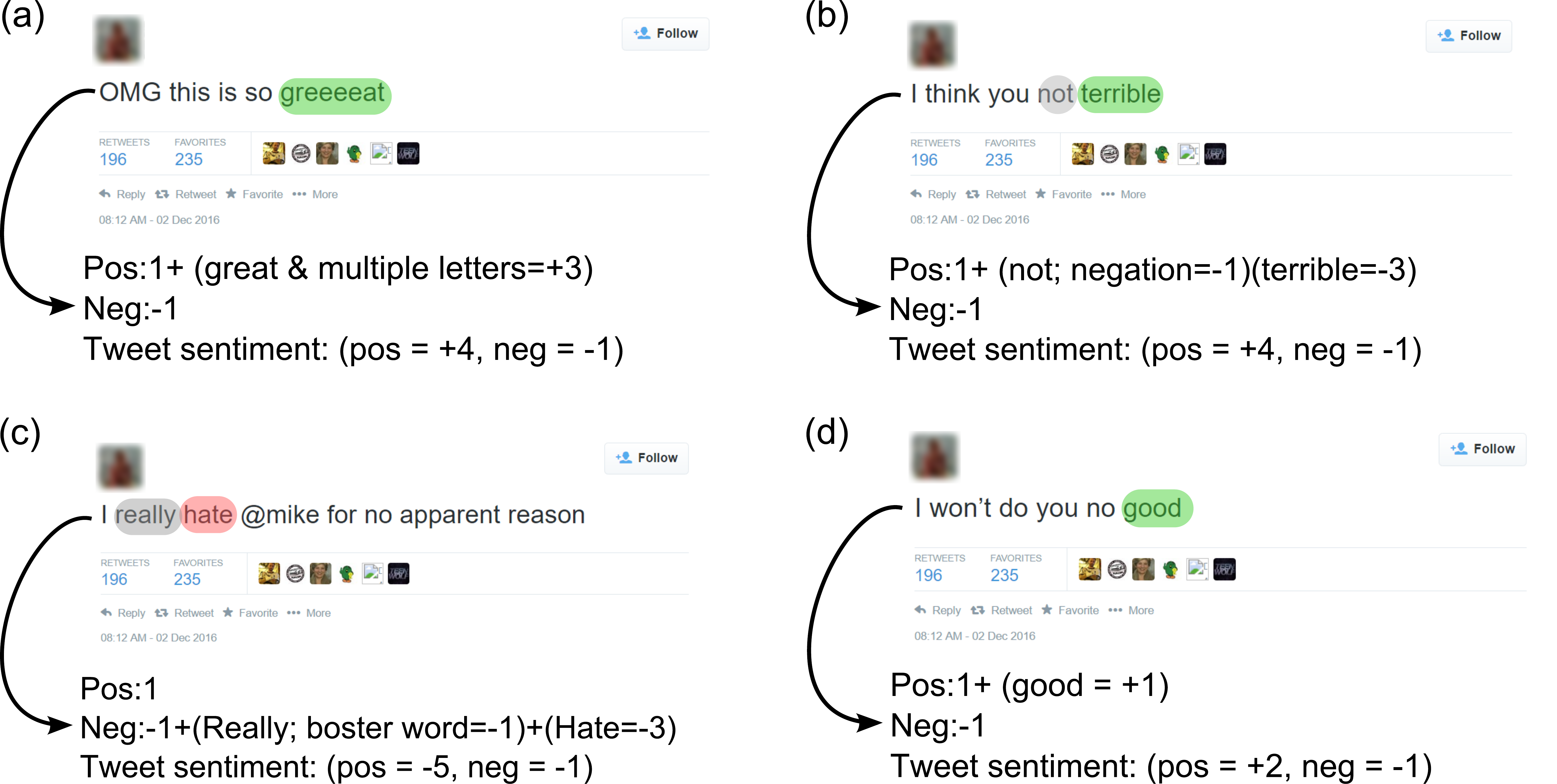}}
  \caption{Examples of how SentiStrength assigns sentiment scores to
    various tweets. Highlighted words are detected as positive
    (green), negative (red), negations and boosters (grey). (a): Example
    of multiple letters. (b): Negation and booster words. (d): Failure to
    detect double negatives.}
  \label{fig:SenExample}
\end{figure}

SentiStrength~\cite{Thelwall2013SentiStrHeart} contains a lexicon of
$2,310$ sentiment-annotated words and word stems (i.e., roots of
words). The system finds the sentiment of a string (more precisely a
sentence, or in this case a tweet) of text by matching each word
against the internal lexicon. The positive and negative score of the
string is the total positive and negative scores from its words
normalised to be between $1$ and $5$ for positive sentiment, and $-1$
and $-5$ for negative.  SentiStrength also accounts for some nuances
of the language by including an extensive rule set that includes
negations, repeated letters (for emphasis), and booster
words~\cite{Thelwall2013SentiStrHeart}. The rules for punctuation do
not apply to our dataset as we removed punctuation as part of the
pre-processing of the text.  Figure~\ref{fig:SenExample} contains
examples of how SentiStrength assigns positive and negative scores to
short strings of text, and how in some cases can miscalculate the
sentiment of a tweet.

\section{Accuracy of the classification of users}
\label{app:accuracy}

The classification of community clusters as either \textit{yes},
\textit{no} or \textit{unaligned} in Sec.~\ref{sec:classAccuracy} was
performed manually by annotating a sample of $20\%$ of users in each
group. The classification of users was based on their profile
description and their tweets with the tracked hashtags, and was
blinded to the community cluster of the user. The profile descriptions
are an indicator of which side users are likely to support, as they
often contain hashtags, words or phrases in support of {\it yes} or
{\it no} (e.g., \#equalitymatters or \#marriagematters can indicate
support for {\it yes} or {\it no}). Tweets from the referendum day
often contain references to having voted or supported yes or no (e.g.,
{\it ``I voted for equality \#voteyes \#marref''}). In the absence of
an overt reference to supporting either side we classified the user
after examining all their tweets in our data. We assigned an {\it
  unaligned} label if the user did not show a discernible leaning
towards the {\it yes} or {\it no} side. Typically, users who were
classified as unaligned either had posted few tweets or their tweets
did not have a discernible leaning (e.g., {\it ``Interesting debate
  taking place now about \#marref''}).

After classifying the tweets from our random sample, we computed the
proportion of {\it yes}, {\it no}, or {\it unaligned} supporters in
each community cluster. The proportion of {\it yes} supporters in
community clusters $CC_1$ and $CC_2$ is $90\%$ and $96\%$; as a
result, we labelled these groups as {\it yes} community clusters.  We
labelled $CC_3$ as a {\it no} community cluster because its proportion
of {\it no} supporters is $71\%$.  To find the accuracy of these
labels we construct a confusion matrix~\cite{brodersen2010balanced}
(Table~\ref{tab:conf_yesno}), which provides a breakdown of true and
false positives.

\begin{table}[tp]
\centering
\begin{tabular}{ll|ll|l}
  & \multicolumn{1}{c|}{} & \multicolumn{2}{c|}{Actual}
  & \multicolumn{1}{l}{} \\
  & & Yes & No & Total  \\ \hline
\multirow{2}{*}{Classification} & Yes & 297 (True yes) & 23 (True no) & 320 \\
 & No  & 29 (False yes) & 9 (False no) & 38 \\ \hline
 & Total & 326 & 32  & 358                 
\end{tabular}
\caption{Confusion matrix with the number of correct and incorrect
  classification for yes and no voters.}
\label{tab:conf_yesno}
\end{table}

We can calculate the overall and balanced accuracy for {\it yes} and
{\it no} supporters using Table~\ref{tab:conf_yesno}. The overall
accuracy is the ratio of ``true yes'' and ``true no'' supporters
($297$ and $23$, respectively) to the total number of users in the
sample ($358$). The overall accuracy for the sample is $89\%$.
However, overall accuracy is known to be biased towards more frequent
classes. To correct for this bias we calculate the balanced
accuracy~\cite{brodersen2010balanced} by calculating the fraction of
correctly classified {\it yes} or {\it no} supporters out of the total
number of actual supporters and averaging the two ($297/326$ and
$23/32$, respectively). The balanced accuracy is then ($0.5(297/326 +
23/32) = 0.81$).

In Sec.~\ref{sec:classAccuracy} we labelled the community clusters in
terms of both the dominant user leaning and activity levels. We
defined $CC_1$ as the {\it Passive yes} community cluster, $CC_2$ the
{\it Active yes}, and $CC_3$ the {\it Active no} community cluster. In
an ideal setting, we would report the balanced accuracy for the three
types of users. In practice, however, it is a difficult and subjective
exercise to discern {\it Passive Yes} from {\it Active Yes} users at
an individual level. The distinction between passive and active is
based on the average user's activity (number of tweets) in each
community cluster shown in Fig.~\ref{fig:TSAverageTPU}(a). Therefore,
a balanced accuracy for both voter classification and activity is not
viable. If we were to make a distinction between
$CC_1$ and $CC_2$ the balanced accuracy would be $1/3(183/205 + 114/121 +
23/32)=0.85$, treating $CC_1$ and $CC_2$ separately.

\section{Robustness of randomisation}\label{app:randomAlternative}

\begin{figure}[t]
  \centering
  \includegraphics[width=0.85\textwidth]{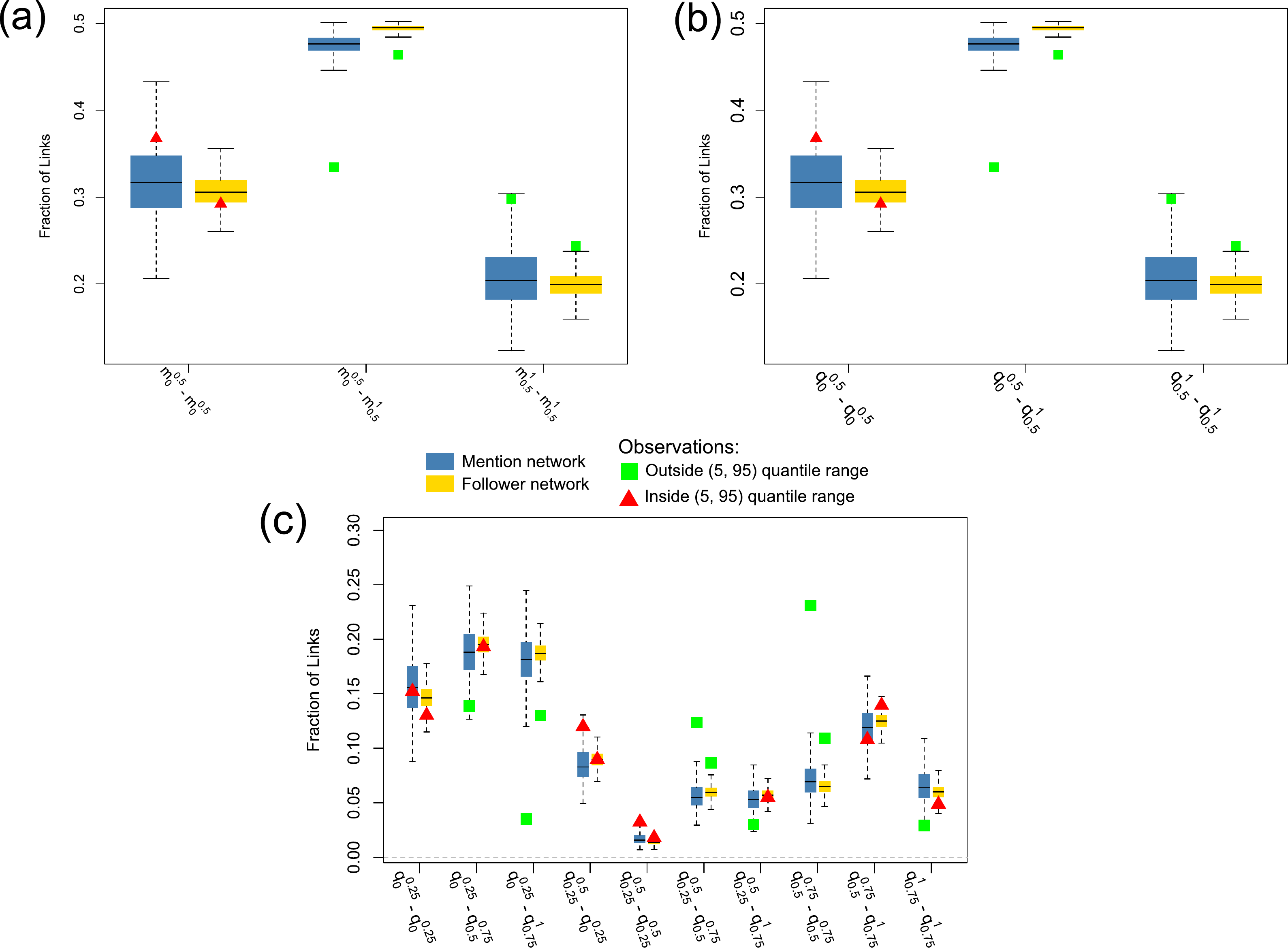}
  \caption{Results after 1,000 randomisation tests in the mention
    network and the follower network. (a): Division of users above and
    below the mean. (b): Division of users above and below the median.
    (c): Division of users into quantiles. Green squares and red
    triangles marks the observed fraction of links in the data. A
    green square indicates that the observed fraction falls outside
    the lower $5\%$ and upper $95\%$ quantiles of the randomised
    distribution. A red square indicates the observed fraction falls
    inside the lower $5\%$ and upper $95\%$ quantiles of the
    randomised distribution.}
        \label{fig:robustres}
\end{figure}

In Sec.~\ref{sec:testingRel} we showed that the sentiment of users'
in-neighbourhoods is positively correlated (in agreement with previous
reports~\cite{Bliss2012}) in both the mention and follower network;
this allows us to use it as a proxy for homophily.  We arrived at this
result by showing that users with similar sentiment, in particular
positive users, were connected more often then we would have expected
by random chance. These results are robust to how we group these users
by sentiment. In Sec.~\ref{sec:testingRel} we applied a coarse
labelling of users according to their sentiment score (``positive'',
``negative'' or ``unknown''). Here we show that a finer labelling of
users also produces similar results. We test three alternative ways of
labelling users:
\begin{enumerate}
  \item Divide users into groups in which the out-sentiment is below
    ($m_0^{0.5}$) and above ($m_{0.5}^1$) the mean.
  \item Divide users into groups in which the out-sentiment is below ($q_{0}^{0.5}$) and
    above ($q_{0.5}^{1}$) the median. 
  \item Divide users into groups by the out-sentiment quartiles ($q_{0}^{0.25}$, $q_{0.25}^{0.5}$,
    $q_{0.5}^{0.75}$, $q_{0.75}^{1}$).
\end{enumerate}
We then randomise the network with these labels in the same way as
described in Sec.~\ref{sec:testingRel}.  Figure~\ref{fig:robustres}
shows the results from the new randomisation tests, which are
consistent with our results in Sec.~\ref{sec:testingRel} of the Main
Text.  The similarity observed between
Fig.~\ref{fig:robustres}(a)~and~(b) is due to the fact that the mean
and median of the out sentiment distribution are close. In both cases
users above the mean and median tend to be connected more than we
would have expected by chance. Figure~\ref{fig:robustres}(c) tells a
similar story, where users in the top two quartiles are more likely to
be connected with each other than what we would expect by chance.

\bibliographystyle{siam}

\end{document}